\documentclass[twocolumn,showpacs,preprintnumbers,amsmath,amssymb]{revtex4}

\usepackage[dvips]{graphicx}

\begin{document}
\title{Lepton flavor violation with supersymmetric Higgs triplets \\
in TeV region for neutrino masses and leptogenesis}

\author{Masato Senami}
\altaffiliation{E-mail address: senami@nucleng.kyoto-u.ac.jp}
\author{Katsuji Yamamoto}
\altaffiliation{E-mail address: yamamoto@nucleng.kyoto-u.ac.jp}
 \affiliation{Department of Nuclear Engineering, Kyoto University,
Kyoto 606-8501, Japan}

\date{\today}

\begin{abstract}
Lepton flavor violating processes
such as $ \mu \to 3e $ and $ \mu \to e \gamma $ are investigated
with supersymmetric Higgs triplet pair $ \Delta $ and $ {\bar \Delta} $
in the light of neutrino masses and experimentally verifiable leptogenesis.
The Higgs triplet mass $ M_\Delta $ is expected to be
in the range of $ 1 - 100 {\rm TeV} $.
The branching ratios of these charged lepton decays
are evaluated in terms of $ M_\Delta $
and the coupling $ f L \Delta L $
of Higgs triplet $ \Delta $ with lepton doublet pairs $ L L $,
which is proportional to the neutrino mass matrix.
They may be reached in the future collider experiments.
In particular, the $ \mu \to 3e $ decay would be observed
indicating the existence of Higgs triplets
with $ M_\Delta \sim 1 - 100 {\rm TeV} $ for $ | f | \sim 0.1 - 1 $,
while the $ \mu \to e \gamma $ decay can be significant
irrespective of $ M_\Delta $ in the supersymmetric model
due to the flavor violation in the slepton mass matrices
induced by the renormalization effects.
\end{abstract}
\maketitle

\section{Introduction}
\label{sec:introduction}

The neutrino masses may be generated naturally
by introducing the electroweak Higgs triplet $ \Delta $
\cite{triplethiggs}.
The effective higher-dimensional operators $ L_i L_j H_u H_u $
of the lepton and Higgs doublets
(with indices $ i , j $ to denote the generations)
are indeed provided by the exchange of Higgs triplet,
as well as the right-handed neutrinos in the usual see-saw mechanism.
The lepton number violation with Higgs triplet
or right-handed neutrinos may further realize
the generation of lepton number asymmetry, leptogenesis,
in the early universe.
Then, the sufficient baryon-to-entropy ratio
can be provided from the lepton number asymmetry
through the electroweak anomalous effect
\cite{FY}.

The leptogenesis has been investigated extensively
in the literature in connection with the neutrino mass generation.
In most scenarios of leptogenesis via lepton number nonconserving decays,
the relevant particles such as right-handed neutrinos and Higgs triplets
are supposed to be much heavier than the electroweak scale.
On the phenomenological point of view, however,
these particles are expected to be alive in the TeV region.
In this respect, it is interesting that the leptogenesis can be realized
with supersymmetric Higgs triplets via multiscalar coherent evolution
after the inflation
\cite{PLB524,PRD66-67}.
While the Higgs triplet mass $ M_\Delta $ was originally
supposed to be in the range $ 10^{9}-10^{14} {\rm GeV} $
\cite{PLB524},
it has been found by reanalizing this leptogenesis scenario
that the successful leptogenesis is possible
even with the Higgs triplet mass
in the TeV region $ M_\Delta \sim 1 - 100 {\rm TeV} $
\cite{newpaper}.
That is, just after the inflation the lepton number asymmetry
appears via multiscalar coherent motion on the flat manifold
of a pair of Higgs triplets $ \Delta $, $ {\bar \Delta} $
and the anti-slepton $ {\tilde e}^c $
in the manner of Affleck-Dine mechanism
\cite{AD,DRT}.
Then, the lepton number asymmetry is fluctuating during some period,
and it is fixed to some significant value
due to the effect of the Higgs triplet mass terms.
It is here essential for fixing the lepton number asymmetry
that the Higgs triplet mass terms should prevail
over the negative thermal log term,
requiring a condition on $ M_\Delta $.
This condition can really be satisfied
for $ M_\Delta \gtrsim 1 {\rm TeV} $
depending on the reheating temperature of the universe
$ T_R < 10^9 {\rm GeV} $
and the mass scale $ M / \lambda \sim 10^{20} - 10^{23} {\rm GeV} $
of the nonrenormalizable superpotential terms for leptogenesis.

If the Higgs triplet mass is $ M_\Delta \sim 1 {\rm TeV} $,
quite interesting phenomenology is provided in the electroweak to TeV region
\cite{raidal,MRS,t3ha,t3hb,t3hc}.
Then, the leptogenesis scenario as well as the neutrino mass generation
with supersymmetric Higgs triplets
can be verified by the future collider experiments.
The Higgs triplets may be discovered by direct production,
and their effects on lepton flavor violation may also be
found in the decays of charged leptons,
$ \mu \rightarrow 3e $, $ \mu \rightarrow e \gamma $, 
$ \tau \rightarrow 3 \mu $, $ \tau \rightarrow \mu \gamma $, and so on.
It is particularly interesting in the Higgs triplet model
that these lepton flavor violating processes are related each other
through the neutrino mass matrix, which is proportional
to the Yukawa coupling $ f_{ij} L_i L_j \Delta $.
The experimental observations on the atmospheric and solar neutrinos
now provide important information about the neutrino masses and mixings
\cite{SK,SNO,CHOOZ,KamLand}.
Then, these relations among the lepton flavor violating processes
in the Higgs triplet model will be tested in the feasible experiments,
as investigated in the literature for the non-supersymmetric model
with $ M_\Delta \sim 100 {\rm GeV} - 1 {\rm TeV} $
\cite{t3ha,t3hb}
and the supersymmetric model
with $ M_\Delta \sim 10^{11} - 10^{14} {\rm GeV} $
through the renormalization effects on the slepton masses
\cite{rossi}.

We here investigate these lepton flavor violating effects
of the supersymmetric Higgs triplets
in the light of neutrino masses and experimentally verifiable leptogenesis.
The Higgs triplet mass $  $ is expected to be
$ M_\Delta \sim 1 - 100 {\rm TeV} $.
This article is organized as follows.
In Sec. \ref{sec:model},
we present the supersymmetric Higgs triplet model,
and describe the neutrino mass generation,
discussing how the Higgs triplets in TeV region can develop naturally
the desired tiny vacuum expectation values.
In Sec. \ref{sec:LFV}, we examine the lepton flavor violating terms
provided with the Higgs triplets, including the renormalization effects.
In Sec. \ref{sec:processes}, we investigate
the lepton flavor violating processes and muon anomalous magnetic moment,
which are related each other through the neutrino mass matrix.
Sec. \ref{sec:sum} is devoted to summary.
The one-loop contributions of supersymmetric Higgs triplets
to the charged lepton radiative decay amplitudes are calculated
in Appendix \ref{Delta-contribution}.

\section{Neutrinos with Higgs triplets}
\label{sec:model}

We investigate an extension of the minimal supersymmetric standard model
by introducing a pair of Higgs triplets $ \Delta $ and $ {\bar \Delta} $,
which are specified in terms of
$ {\rm SU(3)}_C \times {\rm SU(2)}_L \times {\rm U(1)}_Y $
as
\begin{eqnarray}
\Delta & = & \left( \begin{array}{cc}
\Delta^+ / {\sqrt 2} & \Delta^{++} \\
\Delta^0 & - \Delta^+ / {\sqrt 2}
\end{array} \right) \sim ({\bf 1},{\bf 3},1) , \\
{\bar \Delta} & = &
\left( \begin{array}{cc}
{\bar \Delta}^- / {\sqrt 2} & {\bar \Delta}^0 \\
{\bar \Delta}^{--} & - {\bar \Delta}^- / {\sqrt 2}
\end{array} \right) \sim ({\bf 1},{\bf 3},-1) .
\end{eqnarray}
The lepton doublets $ L_i = ( \nu_i , l_i ) $,
anti-lepton singlets $ l_i^c $ ($ i = e , \mu , \tau $)
and the Higgs doublets $ H_u $, $ H_d $ are given as usual.
The generic lepton number conserving superpotential
for the leptons and Higgs fields is given by
\begin{eqnarray}
W_0 = h_{ij} L_i H_d l^c_j + \mu H_u H_d 
    + \frac{1}{\sqrt 2} f_{ij} L_i \Delta L_j
    + M_\Delta {\bar \Delta} \Delta ,
\label{W0}
\end{eqnarray}
where the lepton basis is taken at the electroweak scale $ M_W $
with the diagonal Yukawa coupling $ h $,
and the dilepton coupling is given by a symmetric matrix
$ f = f^{\rm T} $.
The lepton numbers are assigned to the Higgs triplets as
\begin{eqnarray}
Q_L ( \Delta ) = -2 , \ Q_L ( {\bar \Delta} ) = 2 .
\end{eqnarray}
Then, the lepton number violating terms may also be included
in the superpotential as
\begin{eqnarray}
W_{\rm LV} = \xi_1 H_u {\bar \Delta} H_u + \xi_2 H_d \Delta H_d  .
\end{eqnarray}
The Higgs triplets are $ R $-parity even,
and we here do not consider the $ R $-parity violation for definiteness.

The Higgs triplets develop nonzero vacuum expectation values
(VEV's) due to the effects of $ W_{\rm LV} $ as
\begin{eqnarray}
\langle {\Delta}^0 \rangle
= - c_1 \frac{\xi_1 \langle H_u \rangle^2 }{M_\Delta} ,
\langle {\bar \Delta}^0 \rangle
= - c_2 \frac{\xi_2 \langle H_d \rangle^2 }{M_\Delta} .
\label{tripletvev}
\end{eqnarray}
The factors $ c_1 , c_2 \sim 1 $ for $ \xi_1 \sim \xi_2 $
are determined precisely by minimizing the scalar potential
including the soft supersymmetry breaking terms
with the mass scale $ m_0 \sim 10^3 {\rm GeV} $
($ c_1 = c_2 = 1 $ in the limit of $ \mu , m_0 \rightarrow 0 $).
It should be noted here that these VEV's are induced
by the $ \xi_1 $ and $ \xi_2 $ couplings
explicitly violating the lepton number conservation.
Hence, the so-called triplet Majoron does not appear
from the $ \Delta $ and $ {\bar \Delta} $ fields,
which rather acquire masses $ \simeq M_\Delta $.
The slepton fields $ {\tilde L}_i $, $ {\tilde l}^c_i $ do not develop VEV's
since the $ R $-parity is still preserved by the VEV's of Higgs triplets.

The neutrino mass matrix is provided by the VEV of the Higgs triplet as
\begin{equation}
M_{\nu} = f {\sqrt 2} \langle \Delta^0 \rangle ,
\label{Mnu}
\end{equation}
which is diagonalized with a unitary matrix $ U $ as
\begin{equation}
U^{\rm T} M_\nu U = {\rm diag} ( m_1 , m_2 , m_3 ) .
\end{equation}
The charged lepton mass matrix is also given as
\begin{equation}
M_l = h \langle H_d \rangle = {\rm diag} ( m_e , m_\mu , m_\tau ) .
\label{Ml}
\end{equation}
Here, the flavor structure of leptons is described at $ M_W $
by the $ f $ coupling with the diagonal $ h $ coupling.
This neutrino mass matrix (\ref{Mnu}) should reproduce
the masses and mixing angles inferred
from the data of neutrino experiments
\cite{SK,SNO,CHOOZ,KamLand}.
Then, by considering Eqs. (\ref{tripletvev}) and (\ref{Mnu})
with $ m_i \lesssim 10^{-1} {\rm eV} $
a constraint on the magnitude of $ f $ coupling is placed roughly as
\begin{eqnarray}
| f | \lesssim 10^{-1}
\left( \frac{\xi}{10^{-10}} \right)^{-1}
\left( \frac{M_\Delta}{10^3 {\rm GeV}} \right) .
\label{fcond}
\end{eqnarray}
Here, the magnitude of the lepton number violating couplings
is supposed to be very small as $ \xi_1 , \xi_2 \sim \xi \sim 10^{-10} $
for $ M_\Delta \sim 10^3 {\rm GeV} $.

These tiny lepton number violating couplings $ \xi_1 $, $ \xi_2 $
inducing the VEV's of Higgs triplets may be explained as follows
\cite{PLB524}.
Suppose that the lepton number/$ R $-parity violation originates
in the Planck scale physics.
Then, it may be provided
with certain higher-order effective superpotential terms as
\begin{eqnarray}
W_{\rm LV}^\prime
= \xi^\prime_1 \frac{{\bar S} H_u {\bar \Delta} H_u}{M_{\rm P}}
+ \xi^\prime_2 \frac{S H_d \Delta H_d}{M_{\rm P}}
\end{eqnarray}
with the reduced Planck mass
$ M_{\rm P} = m_{\rm P} / {\sqrt{8 \pi}}
= 2.4 \times 10^{18} {\rm GeV} $.
Here, some singlet superfields $ S $ and $ {\bar S} $
of $ R $-parity odd with $ Q_L = 1 , - 1 $, respectively
are also considered.
The lepton number/$ R $-parity violating terms
$ S H_u H_d $ and $ S \Delta {\bar \Delta} $ are hence excluded.
These singlet fields may have the lepton number/$ R $ preserving
superpotential terms,
\begin{eqnarray}
W_S = M_S S {\bar S} + \lambda_S \frac{ S S {\bar S} {\bar S} }{M_{\rm P}},
\label{sym-bre}
\end{eqnarray}
where the Higgs singlet mass is assumed to be $ M_S \sim 10^3 {\rm GeV} $
as well as the Higgs triplet mass $ M_\Delta \sim 10^3 {\rm GeV} $.
Without cubic terms for the Higgs singlets,
they are considered as flatons
\cite{flaton}, and may develop large VEV's
with vanishing $ F $-temrs $ | F_S | , | F_{\bar S} | \approx 0 $ as
\begin{eqnarray}
\langle S \rangle \sim \langle {\bar S} \rangle
\sim \sqrt{M_S M_{\rm P}} \sim 10^{10} {\rm GeV} .
\end{eqnarray}
Then, the lepton number violating couplings
$ \xi_1 $ and $ \xi_2 $ are derived effectively as
\begin{eqnarray}
\xi_1 = \xi_1^\prime ( \langle {\bar S} \rangle / M_{\rm P} ) ,
\xi_2 = \xi_2^\prime ( \langle S \rangle / M_{\rm P} )
\label{LFVcoupling}
\end{eqnarray}
with the tiny factor desired for $ \xi $ in Eq. (\ref{fcond}),
\begin{equation}
\langle S \rangle / M_{\rm P} , \langle {\bar S} \rangle / M_{\rm P}
\sim \sqrt{M_S / M_{\rm P}} \sim 10^{-8} .
\end{equation}

It is also notable that the smallness of the Higgs triplet VEV's
may be explained elegantly in the context of large extra dimensions
\cite{MRS}.

\section{Lepton flavor violation}
\label{sec:LFV}

We here examine the lepton flavor violating couplings
provided with Higgs triplets,
including the renormalization group effects.

\subsection{Yukawa couplings}

The lepton basis is taken
with the diagonal Yukawa coupling $ h $ at $ M_W $
in Eqs. (\ref{W0}) and (\ref{Ml}):
\begin{equation}
h_{ij} = h_i \delta_{ij} .
\end{equation}
(Henceforth $ M_W $ is omitted for the quantities at the electroweak scale.)
Then, the lepton flavor violation, which is provided by the $ f $ coupling
at $ M_W $, is linked directly to the neutrino mass matrix,
as seen in Eq. (\ref{Mnu})
\cite{t3ha,t3hb,rossi}.
This is a very interesting feature of Higgs triplet model.
Specifically, the $ f $ coupling is given
in terms of the neutrino masses ($ m_i $),
mixing angles ($ \theta_{ij} $)
and $ CP $ violating phases ($ \delta , \alpha_1 , \alpha_2 $) as
\begin{equation}
f_{ij} = | f | \sum_k U_{ik}^* U_{jk}^* ( m_k / m_{\rm atm} ) ,
\label{f_def}
\end{equation}
where $ m_{\rm atm} \equiv \sqrt{\Delta m_{\rm atm}^2} $
with $ \Delta m_{\rm atm}^2 \sim 3 \times 10^{-3} {\rm eV}^2 $.
The explicit form of the Maki-Nakagawa-Sakata (MNS) matrix
$ U $ (lepton mixing matrix) \cite{MNS}
is given in a review of the Particle Data \cite{PDG}.
The mean magnitude of the $ f $ coupling is given suitably by
\begin{eqnarray}
| f | \equiv m_{\rm atm} / ( {\sqrt 2} \langle \Delta^0 \rangle ) ,
\end{eqnarray}
which is constrained, as seen in Eq. (\ref{fcond}),
with $ \langle \Delta^0 \rangle $ in terms of $ \xi $ and $ M_\Delta $.

The flavor violation appears in the $ h $ coupling
at certain unification scale $ M_G $ such as
the grand unification or gravitational scale
through the renormalization effects.
In the bottom-up view point $ M_W \rightarrow M_G $,
the relevant couplings at $ M_G $ are evaluated with those at $ M_W $ as
\begin{eqnarray}
h_{ij} ( M_G ) &=& c_{h i} h_i \delta_{ij} + ( \Delta_f h )_{ij} ,
\\
f_{ij} ( M_G ) &=& c_{f ij} f_{ij} + ( \Delta_f f )_{ij} ,
\end{eqnarray}
where the sum is not taken over $ i , j $.
The factors $ c_{h i} , c_{f ij} \sim 1 $
are provided by the gauge and $ h $ couplings.
The remaining terms provided by the $ f $ coupling
are calculated in the leading-log approximation as
\begin{eqnarray}
( \Delta_f h )_{ij}
& \simeq & (3/2) h_i ( f^\dagger f )_{ij} t_G ,
\\
( \Delta_f f )_{ij}
& \simeq & 3 ( f f^\dagger f )_{ij} t_G ,
\end{eqnarray}
where
\begin{equation}
t_G \equiv ( 1 / 8 \pi^2 ) {\rm ln} ( M_G / M_W ) \sim 0.4 .
\end{equation}

\subsection{Slepton mass terms}

The flavor violation also appears in the soft supersymmetry breaking terms.
We may assume the universality of the soft supersymmetry breaking terms
at the unification scale $ M_G $, i.e.,
the soft masses of scalar fields are given by the common mass
$ m_0 $, and the $ A $-terms are given by $ a_0 m_0 $
with $ a_0 \sim 1 $.
Then, in the top-down view point $ M_G \rightarrow M_W $
the soft mass terms at $ M_W $ are calculated particularly
for the left-handed slepton doublets $ {\tilde L} $
and the right-handed charged slepton singlets $ {\tilde l}^c $
\cite{rossi} as
\begin{eqnarray}
( M^2_{\tilde L} )_{ij}
& = & c_{\tilde L} m_0^2 \delta_{ij}
+ ( \Delta_{h+f} M^2_{\tilde L} )_{ij} ,
\\
( M^2_{{\tilde l}^c} )_{ij}
& = & c_{{\tilde l}^c} m_0^2 \delta_{ij}
+ ( \Delta_h M^2_{{\tilde l}^c} )_{ij} .
\end{eqnarray}
Here, the contributions of the gauge couplings
are included in the factors
$ c_{\tilde L} , c_{{\tilde l}^c} \sim 1 $, and
those of the $ h $ and $ f $ couplings are given by
\begin{eqnarray}
\Delta_{h+f} M^2_{\tilde L}
& \simeq & - m_0^2 [ ( 3 + a_0^2 ) h^\dagger ( M_G ) h ( M_G )
\nonumber \\
&& - ( 9 + 3 a_0^2 ) f^\dagger ( M_G ) f ( M_G ) ] t_G ,
\\
\Delta_h M^2_{{\tilde l}^c}
& \simeq & - m_0^2 ( 6 + 2 a_0^2 ) h^\dagger ( M_G ) h ( M_G ) t_G .
\label{sleptonmass}
\end{eqnarray}
The $ A_h $ term of the $ h $ coupling is also given at $ M_W $ by
\begin{equation}
( A_h )_{ij} = a_h m_0 h_{ij} ( M_G ) + ( \Delta A_h )_{ij}
\end{equation}
with $ a_h \sim a_0 $ including the effects of gauge couplings and
\begin{eqnarray}
\Delta A_h
& \simeq & - (9/2) a_0 m_0 h ( M_G ) [ h^\dagger ( M_G ) h ( M_G )
\nonumber \\
&& + f^\dagger ( M_G ) f ( M_G ) ] t_G .
\end{eqnarray}

The charged slepton mass matrix is given
in the basis of $ ( {\tilde l} , {\tilde l}^{c *} ) $ by
\begin{equation}
{\cal M}^2_{\tilde l} = \left( \begin{array}{cc}
M^2_{{\tilde l}LL} & M^2_{{\tilde l}LR} \\
{} & {} \\
M^2_{{\tilde l}RL} & M^2_{{\tilde l}RR} \end{array} \right) ,
\label{lmass}
\end{equation}
where the submatrices are given by
\begin{eqnarray}
M^2_{{\tilde l}LL} & = &  M^2_{\tilde L} + M_l^2 ,
\\
M^2_{{\tilde l}RR} & = & M^2_{{\tilde l}^c} + M_l^2 ,
\\
M^2_{{\tilde l}LR} & = & M^{2 \dagger}_{{\tilde l}RL}
= \langle H_d \rangle A_h + \tan \beta \mu M_l ,
\end{eqnarray}
with $ \tan \beta \equiv \langle H_u \rangle / \langle H_d \rangle $.
The sneutrino mass matrix is also given by
\begin{equation}
{\cal M}^2_{\tilde \nu} = M^2_{\tilde L} ,
\label{numass}
\end{equation}
where the tiny lepton number violating term
related to the Majorana neutrino mass matrix is neglected
in a good approximation.
The flavor changing components are particularly
calculated in the leading order of $ | f |^2 $ as
\begin{align}
\frac{( M^2_{{\tilde l}LL} )_{ij}|_{i \neq j}}{m_0^2}
&\simeq - 3 ( 3 + a_0^2 )( 1 + c_{hi} h_i^2 t_G ) t_G ( f^\dagger f )_{ij} ,
\label{MLL} \\
\frac{( M^2_{{\tilde l}RR} )_{ij}|_{i \neq j}}{m_0^2}
&\simeq - 6 ( 3 + a_0^2 ) c_{hi} h_i^2 t_G^2 ( f^\dagger f )_{ij} ,
\label{MRR} \\
\frac{( M^2_{{\tilde l}LR} )_{ij}|_{i \neq j}}{m_0^2}
&\simeq - \frac{9}{2} a_0 \frac{m_{l_i}}{m_0} ( 1 + 3 c_{hi} h_i^2 t_G ) t_G
( f^\dagger f )_{ij}  ,
\label{MLR} \\
\frac{M^2_{\tilde \nu}|_{i \neq j}}{m_0^2}
&\simeq - 3 ( 3 + a_0^2 )( 1 + c_{hi} h_i^2 t_G ) t_G ( f^\dagger f )_{ij} ,
\label{MSNU}
\end{align}
where the values of $ f $ and $ h $ couplings are taken at $ M_W $.
It is noticed that these leading contributions of flavor violation
are determined essentially by $ t_G ( f^\dagger f )_{ij} $ ($ i \neq j $)
\cite{rossi}
with the significant log factor $ t_G \sim 0.4 $ in the present scheme
of $ M_\Delta \sim 10^3 {\rm GeV} $.

\section{Charged lepton processes }
\label{sec:processes}

We investigate the charged lepton processes in order,
to which the supersymmetric Higgs triplets in TeV region
may provide significant contributions.
Such effects are expected to show the evidence of Higgs triplets
particularly related to the neutrino masses and mixings.

\subsection{$\mu \to 3 e $ and $ \tau \to 3 \mu $}

The leading contribution to the $ \mu \to  3 e $ decay
is provided at the tree level mediated by the Higgs triplet.
The supersymmetric contributions, on the other hand,
appear at the one-loop level through the flavor violation
in the slepton sectors \cite{HMTYY}.
They are, however, negligible compared
to the tree-level contribution for $ M_\Delta \sim 10^3 {\rm GeV} $.
The branching ratio is calculated
\cite{swartz} as
\begin{eqnarray}
&& {\rm Br}(\mu \rightarrow 3 e )
= \frac{| f_{ee}^* f_{\mu e} |^2}{8 g^4}
\left( \frac{M_W}{m_\Delta} \right)^4
\nonumber \\
&& \ \ \ \ \ = 3 \times 10^{-13}
\left( \frac{1 \rm TeV}{m_\Delta} \right)^4
\left( \frac{| I_{\mu \to 3e} |}{0.01} \right)^2
\left( \frac{| f |}{0.1} \right)^4 ,
\nonumber \\
\label{muto3e}
\end{eqnarray}
where
\begin{equation}
f_{ee}^* f_{\mu e} \equiv I_{\mu \to 3e} | f |^2 ,
\end{equation}
and the mass of scalar Higgs triplet is given
including the contribution of soft supersymmetry breaking
($ c_\Delta \sim 1 $) by
\begin{equation}
m_\Delta = {\sqrt{M_\Delta^2 + c_\Delta m_0^2}} .
\label{mDelta}
\end{equation}
The experimental bound is, on the other hand, placed as
$ {\rm Br}( \mu \rightarrow 3 e ) < 1.0 \times 10^{-12} $
\cite{SINDRUM}.
The flavor changing factor $ | I_{\mu \to 3e} | = 0.01 $
is taken in Eq. (\ref{muto3e}) as a reference value.
Its value is evaluated precisely from Eq. (\ref{f_def})
with the neutrino masses and MNS matrix $ U $,
which are inferred from the data of neutrino experiments
\cite{SK,SNO,CHOOZ,KamLand}.
Numerically, we have
$ | I_{\mu \to 3e} | \lesssim 0.03 $ (HI),
$ \lesssim 0.06 $ (DG),
and $ \lesssim 0.2 $ (IH), respectively,
for the hierarchical (HI) case $ m_1 \ll m_2 \ll m_3 $,
the degenerate (DG) case $ m_1 \sim m_2 \sim m_3 $,
and the inverted-hierarchical (IH) case $ m_1 \sim m_2 \gg m_3 $.

\begin{figure}[t]
\begin{center}
\scalebox{.43}{\includegraphics*[1.1cm,14cm][19.4cm,25.7cm]{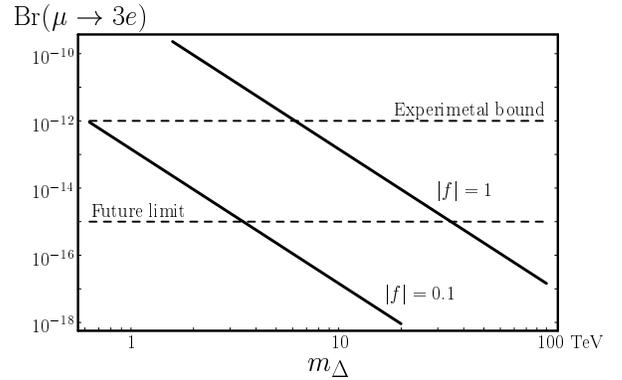}}
\caption{A typical estimate of the branching ratio of $ \mu \to 3 e $
is shown depending on the Higgs triplet mass $ m_\Delta $
for $ | f | = 1 $ and $ | f | = 0.1 $.
}
\label{mu3e}
\end{center}
\end{figure}

A detailed estimate of $ {\rm Br} ( \mu \to 3 e ) $ is presented
in Fig. \ref{mu3e} depending on the Higgs triplet mass $ m_\Delta $.
Typical values of the neutrino masses and mixings are taken
in the HI case as
\begin{eqnarray}
&& ( m_1 , m_2 , m_3 )
= ( 10^{-3} {\rm eV} , 8 \times 10^{-3} {\rm eV} ,
5 \times 10^{-2} {\rm eV} ) ,
\nonumber \\
&& ( \sin \theta_{12} , \sin \theta_{23} , \sin \theta_{13} )
= ( 1/2 , 1/{\sqrt 2} , 0.1 ) ,
\nonumber
\end{eqnarray}
and the zero $ CP $ violating phases, which provides
\begin{eqnarray}
&& I_{\mu \to 3e} = 0.7 \times 10^{-2} .
\nonumber
\end{eqnarray}
The upper and lower solid lines represent
the results for $ | f | = 1 $ and $ | f | = 0.1 $, respectively.
The present experimental bound $ 1.0 \times 10^{-12} $
and a future sensitivity $ \sim 10^{-15} $
achieved by proposed experiments
\cite{KuOk} are also shown with the upper and lower dashed lines,
respectively.
It is interesting here that through the $ \mu \to 3 e $ decay
the evidence of Higgs triplets may be seen
up to the mass $ M_\Delta \simeq m_\Delta = 100 {\rm TeV} $
for $ | f | \sim 1 $.
This will be promising especially
for obtaining the experimental evidence of leptogenesis in TeV region
with the supersymmetric Higgs triplets.
On the other hand, as discussed later,
the Higgs triplet contributions to the $ \mu \to e \gamma $ decay
are significant even for $ M_\Delta \gg 100 {\rm TeV} $
through renormalization effects.

The branching ratio of $ \tau \to 3 \mu $ is also estimated as
\begin{equation}
{\rm Br} ( \tau \rightarrow 3 \mu ) = 2 \times 10^{-11}
\left( \frac{1 \rm  TeV}{m_\Delta} \right)^4
\left( \frac{| I_{\tau \rightarrow 3 \mu} |}{0.2} \right)^2
\left( \frac{| f |}{0.1} \right)^4 ,
\label{tauto3mu}
\end{equation}
where
\begin{equation}
f_{\mu \mu}^* f_{\tau \mu}
\equiv I_{\tau \rightarrow 3 \mu} | f |^2 .
\end{equation}
We have numerically
$ | I_{\tau \rightarrow 3 \mu} | \simeq 0.1 - 0.3 $ (HI),
$ \simeq 0.1 - 0.2 $ (DG), and $ \simeq 0.1 - 0.3 $ (IH), respectively.
This Higgs triplet contribution to the $ \tau \rightarrow 3 \mu $ decay
is far below the experimental bound
$ {\rm Br} ( \tau \rightarrow 3 \mu ) < 3.8 \times 10^{-7} $
\cite{Belle-1}
for $ m_\Delta \sim 1 {\rm TeV} $ and $ | f | \lesssim 0.1 $.
Similar estimates are made for the leptonic three-body decays,
$ \tau \rightarrow {\bar e} \mu \mu $, and so on
\cite{t3hb}.

\subsection{$ \mu \to e \gamma $ and $ \tau \to \mu \gamma $}

The flavor changing radiative decays
such as $ \mu \to e \gamma $ and $ \tau \to \mu \gamma $
are induced by the one-loop diagrams.
In the Higgs triplet model,
the non-supersymmetric contribution is given
by the $ L $-$ \Delta $ loop,
which is almost independent of the mass of the internal lepton
for $ m_l \ll m_\Delta $ \cite{t3ha,t3hb}.
The supersymmetric partner of this contribution is given
by the $ {\tilde L} $-$ {\tilde \Delta} $ loop.
The flavor violation appears even in the internal slepton line
through the renormalization effects,
though its contribution is sufficiently small
for $ | f | \lesssim 10^{-1} $.
The flavor violation in the slepton mass matrices
also provides the supersymmetric contributions
of the $ {\tilde l} $-$ {\tilde \chi}^0 $ (neutralino) loop
and the $ {\tilde \nu} $-$ {\tilde \chi}^- $ (chargino) loop,
as in the minimal supersymmetric standard model
\cite{HMTYY}.
For the case of very large Higgs triplet mass
such as $ M_\Delta \sim 10^{11} - 10^{14} {\rm GeV} $,
the $ {\tilde l} $-$ {\tilde \chi}^0 $
and $ {\tilde \nu} $-$ {\tilde \chi}^- $ contributions are dominant
\cite{rossi},
while the $ L $-$ \Delta $ and $ {\tilde L} $-$ {\tilde \Delta} $
contributions are negligible
due to the suppression factor $ ( m_0 / M_\Delta )^2 $.
On the other hand, for the case of $ M_\Delta \sim m_0 \sim 1 {\rm TeV} $,
as motivated for the direct detection of Higgs triplet,
these contributions may be comparable.

In this interesting case of $ M_\Delta \sim m_0 \sim 1 {\rm TeV} $,
we investigate the charged lepton radiative decays
and their intimate relation to the leptonic three-body decays
of charged leptons through the neutrino mass matrix
proportional to the $ f $ coupling.
In particular, the supersymmetric contributions
of the $ {\tilde l} $-$ {\tilde \chi}^0 $
and $ {\tilde \nu} $-$ {\tilde \chi}^- $ loops
may become most significant for certain range of the model parameters,
while those of the $ {\tilde L} $-$ {\tilde \Delta} $ loop
are comparable to or even larger than
their non-supersymmetric partners of the $ L$-$ \Delta $ loop
for $ M_{\tilde \Delta} = M_\Delta < m_\Delta $.
Then, the relations between the decays $ \mu \rightarrow 3e $, etc.
and the decays $ \mu \rightarrow e \gamma $, etc.,
as found in the non-supersymmetric case
\cite{t3ha,t3hb},
may be modified to some extent,
since the radiative decays are enhanced
due to the supersymmetric contributions
\cite{rossi} with the log-factor $ t_G \sim 0.4 $.

We now estimate the branching ratio of $ \mu \to e \gamma $ decay.
The decay amplitude is generally given by
\begin{equation}
T ( \mu \to e \gamma ) = e \epsilon^{\alpha *} \bar{u}_e
 \left[ i \sigma_{\alpha \beta} q^\beta (A_L P_L + A_R P_R) \right] u_\mu .
\end{equation}
Then, the decay rate is given by
\begin{eqnarray}
\Gamma ( \mu \to e \gamma )
= \frac{ e^2 }{ 16 \pi } m_\mu^3 ( | A_L |^2 + | A_R |^2 ) ,
\end{eqnarray}
and the branching ratio is calculated by
\begin{equation}
{\rm Br}(\mu \rightarrow e \gamma)
= \frac{\Gamma ( \mu \to e \gamma )} {G_F^2 m_\mu^5 / 192 \pi^3} .
\end{equation}

The left-handed and right-handed decay amplitudes are calculated
in the leading order by combining the one-loop contributions:
\begin{eqnarray}
A_{L,R} = A_{L,R}^{{\tilde \chi}^0} + A_{L,R}^{{\tilde \chi}^-}
 + A_{L,R}^\Delta + A_{L,R}^{\tilde \Delta} .
\end{eqnarray}
The formulas for calculating the contributions
$ A_{L,R}^{{\tilde \chi}^0} $ and $ A_{L,R}^{{\tilde \chi}^-} $
of the neutralinos and charginos are presented in the literature
\cite{HMTYY}.
The contributions $ A_{L,R}^\Delta $ and $ A_{L,R}^{\tilde \Delta} $
of the supersymmetric Higgs triplets
are calculated in Appendix \ref{Delta-contribution}.
Then, the decay amplitudes are given specifically as
\begin{eqnarray}
A_L & = &
\frac{m_\mu}{32 \pi^2}
I_{\mu \to e\gamma} | f |^2
\left[ \frac{G_L^{\tilde \chi}}{m_0^2}
+ \frac{G_L^\Delta}{m_\Delta^2} 
+ \frac{G_L^{\tilde \Delta}}{M_\Delta^2} \right] ,
\label{amplL}
\\
A_R & = &
\frac{m_\mu}{32 \pi^2}
I_{\mu \to e\gamma} | f |^2
\left[ \frac{G_R^{\tilde \chi}}{m_0^2}
+ \frac{G_R^\Delta}{m_\Delta^2} 
+ \frac{G_R^{\tilde \Delta}}{M_\Delta^2} \right] ,
\label{amplR}
\end{eqnarray}
where
\begin{eqnarray}
\sum_k f_{ek}^* f_{\mu k}
\equiv I_{\mu \to e\gamma} | f |^2 .
\label{Imueg}
\end{eqnarray}

These leading contributions to the decay amplitudes
are proportional to the flavor changing factor
$ ( f^\dagger f )_{e \mu} = \sum_k f_{ek}^* f_{\mu k} $ ($ f = f^{\rm T} $),
as seen in Eq. (\ref{Imueg}).
This is realized for the $ {\tilde l} $-$ {\tilde \chi}^0 $
and $ {\tilde \nu} $-$ {\tilde \chi}^- $ loops
by using the mass-insertion method with the flavor changing elements
of slepton mass matrices in Eqs. (\ref{MLL}) -- (\ref{MSNU}).
As for the $ L $-$ \Delta $ and $ {\tilde L} $-$ {\tilde \Delta} $ loops,
the flavor-dependence of the masses of intermediate states
can be neglected in a good approximation.
Then, the factor $ ( f^\dagger f )_{e \mu} $ is extracted
from the two vertices in the loop diagram.
(See also Appendix \ref{Delta-contribution} for the detail.)
It should, however, be remarked that the contribution
of $ k = \tau $ in Eq. (\ref{Imueg})
may be modified to some extent for $ \tan \beta \gtrsim 30 $.
This is because the renormalization effects
on the $ {\tilde \tau} $ and $ {\tilde \nu}_\tau $ masses
by the Yukawa coupling $ h_\tau $ becomes significant
especially for the $ {\tilde \tau} $-$ {\tilde \Delta}^{++} $
and $ {\tilde \nu}_\tau $-$ {\tilde \Delta}^+ $ loops.
Furthermore, the renormalization effects may modify
significantly the flavor structure of these amplitudes
for the large $ f $ coupling as $ | f | \sim 0.5 - 1 $.
At present, there is no strong motivation to pursue such special cases.

As a typical example, the factors $ G_{L,R}^{\tilde \chi} $,
$ G_{L,R}^\Delta $ and $ G_{L,R}^{\tilde \Delta} $
are evaluated numerically as
\begin{eqnarray}
&& G_L^{\tilde \chi} = 0.20 , \
G_L^\Delta = 0.8 \times 10^{-3} , \
G_L^{\tilde \Delta} = 1.0 \times 10^{-3} ,
\nonumber \\
&& G_R^{\tilde \chi} = 1.35 , \
G_R^\Delta = 0.17 , \
G_R^{\tilde \Delta} = 0.21 ,
\label{Gs}
\end{eqnarray}
by taking the parameters as
$ M_{\tilde \Delta} = M_\Delta = 700 {\rm GeV} $,
$ m_\Delta = 1000 {\rm GeV} $,
$ m_0 = 700 {\rm GeV} $, $ a_0 = 1 $,
$ \tan \beta = 3 $, $ \mu = 1000 {\rm GeV} $,
$ M_1 = 300 {\rm GeV} $ and $ M_2 = 600 {\rm GeV} $
($ M_1 $ and $ M_2 $ are the gaugino masses
of $ {\rm U(1)}_Y $ and $ {\rm SU(2)}_L $, respectively).
Here, $ G_R^{\tilde \chi} $ is somewhat enhanced by $ \tan \beta $
coming from the $ {\tilde \mu}_L $-$ {\tilde \mu}_R $ flip
with $ ( M^2_{{\tilde l}LR} )_{\mu \mu} $
and the $ \mu_R $-$ {\tilde \mu}_L $-$ {\tilde H}_d^0 $ vertex
\cite{HMTYY,rossi}.
The small ratio $ G_L^\Delta / G_R^\Delta
= G_L^{\tilde \Delta} / G_R^{\tilde \Delta} = m_e / m_\mu $
is attributed to the chirality flip of the external charged leptons.
Then, the branching ratio is estimated as
\begin{eqnarray}
{\rm Br}( \mu \rightarrow e \gamma )
&=& 7 \times 10^{-12} \left( \frac{G}{3} \right)^2
\nonumber \\
& \times & \left( \frac{1 \rm TeV}{m_\Delta} \right)^4
\left( \frac{| I_{\mu \to e \gamma} |}{0.1} \right)^2
\left( \frac{| f |}{0.1} \right)^4 ,
\label{mutoe}
\end{eqnarray}
which should be compared to the experimental bound
$ {\rm Br}( \mu \rightarrow e \gamma ) < 1.2 \times 10^{-11} $
\cite{PDG}.
Here, we take $ G = 3 $ as a reference value for
\begin{equation}
G \equiv
\left( \sum_{K=L,R} \left| r_{\tilde \chi} G_K^{\tilde \chi}
+ G_K^\Delta + r_{\tilde \Delta} G_K^{\tilde \Delta} \right|^2 \right)^{1/2}
\end{equation}
with $ r_{\tilde \chi} \equiv ( m_\Delta / m_0 )^2 $
and $ r_{\tilde \Delta} \equiv ( m_\Delta / M_\Delta )^2 $.
This net $ G $ factor is actually calculated
depending on the various parameters, as seen from Eq. (\ref{Gs}).
It is usually of $ O(1) $ for the reasonable parameter range.
The weights of supersymmetric contributions are relatively enhanced
in $ G $ due to $ r_{\tilde \chi} , r_{\tilde \Delta} > 1 $
for $ m_\Delta > m_0 , M_\Delta $ from Eq. (\ref{mDelta}),
compared to the non-supersymmetric ones.
We have also numerically
$ | I_{\mu \rightarrow e \gamma} | \lesssim 0.2 $ (HI),
$ \lesssim 0.1 $ (DG), and $ \lesssim 0.2 $ (IH), respectively.
This expected branching ratio $ {\rm Br}( \mu \rightarrow e \gamma ) $
really becomes larger by one order or so
due to the supersymmetric contributions
than that of the non-supersymmetric case
\cite{t3ha,t3hb}.
It should also be remarked that the $ \mu \rightarrow e \gamma $ decay
can be a good test to distinguish the supersymmetric Higgs triplets
from the non-supersymmetric ones.
This is because in the non-supersymmetric model
the left-handed decay amplitude $ A_L = A_L^\Delta $
is much smaller than the right-handed one $ A_R = A_R^\Delta $
due to the suppression with $ m_e / m_\mu $.

We can make a similar estimate on the branching ratio
of $ \tau \to \mu \gamma $ as
\begin{eqnarray}
{\rm Br} ( \tau \rightarrow \mu \gamma )
&=& 3 \times 10^{-11} \left( \frac{G}{3} \right)^2
\nonumber \\
& \times & \left( \frac{1 \rm  TeV}{m_\Delta} \right)^4
\left( \frac{| I_{\tau \to \mu \gamma} |}{0.5} \right)^2
\left( \frac{| f |}{0.1} \right)^4 ,
\label{tautomu}
\end{eqnarray}
where
\begin{equation}
\sum_k f_{\mu k}^* f_{\tau k}
\equiv I_{\tau \to \mu \gamma} | f |^2 .
\end{equation}
We have numerically
$ | I_{\tau \rightarrow \mu \gamma} | \simeq 0.4 - 0.5 $ (HI),
$ \simeq 0.1 - 0.4 $ (DG), and $ \simeq 0.4 - 0.5 $ (IH), respectively.
This Higgs triplet contribution to the $ \tau \rightarrow \mu \gamma $ decay
is much smaller than the experimental bound
$ {\rm Br} ( \tau \rightarrow \mu \gamma ) < 3.1 \times 10^{-7} $
\cite{Belle-2}
for $ m_\Delta \sim 1 {\rm TeV} $ and $ | f | \lesssim 0.1 $.

\subsection{Muon anomalous magnetic moment}

The contributions of the $ f $ coupling
to the muon anomalous magnetic moment
mainly appear through the $ \Delta $-$ L $
and $ {\tilde \Delta} $-$ {\tilde L} $ loops.
The magnitude of these contributions
are estimated roughly for $ M_\Delta \sim m_0 $ as
\begin{eqnarray}
| \Delta_f a_\mu | & \sim &
\frac{1}{8 \pi^2} \left( \frac{m_\mu}{m_\Delta} \right)^2
\sum_k | f_{\mu k} |^2
\nonumber \\
& \sim & 10^{-12} 
\left( \frac{1 \rm TeV}{m_\Delta} \right)^2
\left( \frac{| f |}{0.1} \right)^2 .
\end{eqnarray}
Hence, the contributions of the $ f $ coupling
to the muon anomalous magnetic moment
are found to be harmlessly small.

\section{Summary}
\label{sec:sum}

We have investigated the lepton flavor violating processes
such as $ \mu \to 3e $ and $ \mu \to e \gamma $
with the supersymmetric Higgs triplets
in the light of neutrino masses and experimentally verifiable leptogenesis.
The Higgs triplet mass $ M_\Delta $ is expected to be
in the range of $ 1 - 100 {\rm TeV} $.
The branching ratios of these charged lepton decays
are evaluated in terms of $ M_\Delta $
and the coupling $ f L \Delta L $
of Higgs triplet $ \Delta $ with lepton doublet pairs $ L L $,
which is proportional to the neutrino mass matrix.
They may be reached in the future collider experiments.
In particular, the $ \mu \to 3e $ decay would be observed
indicating the existence of Higgs triplets
with $ M_\Delta \sim 1 - 100 {\rm TeV} $ for $ | f | \sim 0.1 - 1 $,
while $ {\rm Br} ( \mu \to e \gamma ) $ can be significant
irrespective of $ M_\Delta $ in the supersymmetric model 
due to the flavor violation in the slepton mass matrices
induced by the renormalization effects.

\acknowledgments

This work is supported in part by
Grant-in-Aid for Scientific Research on Priority Areas B (No. 13135214)
from the Ministry of Education, Culture, Sports, Science and Technology,
Japan.

\appendix
\section{One-loop contributions of supersymmetric $ \Delta $
to $ l_j \to l_i + \gamma $}
\label{Delta-contribution}

We here present the formulas for calculating
the one-loop contributions of supersymmetric Higgs triplets
to the decay amplitudes of $ l_j \to l_i + \gamma $.

The charged slepton mass eigenstates are determined
by diagonalizing the mass matrix $ {\cal M}^2_{\tilde l} $
in Eq. (\ref{lmass}) with a unitary matrix $ U^{\tilde l} $:
\begin{eqnarray}
{\tilde l}_a = U^{\tilde l}_{ai} {\tilde l}_{Li}
+ U^{\tilde l}_{ai+3} {\tilde l}_{Ri} \ ( a = 1 - 6 ) ,
\label{la}
\end{eqnarray}
where $ {\tilde l}_L \equiv \tilde l $ and
$ {\tilde l}_R \equiv {\tilde l}^{c*} $.
The sneutrino mass eigenstates are determined
by diagonalizing the mass matrix $ {\cal M}^2_{\tilde \nu} $
in Eq. (\ref{numass}) with a unitary matrix $ U^{\tilde \nu} $:
\begin{eqnarray}
{\tilde \nu}_b = U^{\tilde \nu}_{bi} {\tilde \nu}_{Li} \ ( b = 1 - 3 ) ,
\label{nub}
\end{eqnarray}
where $ {\tilde \nu}_{L_i} \equiv {\tilde \nu}_i $.
The interactions of bileptons with scalar Higgs triplet are given
from Eq. (\ref{W0}) by
\begin{eqnarray}
{\cal L}_{\Delta}
&=& - \frac{1}{\sqrt{2}} f_{ij} {\bar l}_i^c P_L l_j \Delta^{++}
- \frac{1}{2} f_{ij} {\bar l}_i^c P_L \nu_j \Delta^+
\nonumber \\
&& 
- \frac{1}{2} f_{ij} {\bar \nu}_i^c P_L l_j \Delta^+
+ \frac{1}{\sqrt{2}} f_{ij} {\bar \nu}_i^c P_L \nu_j \Delta^0
\nonumber \\
&& + {\rm H.c.} .
\label{nonsusy-f}
\end{eqnarray}
The interactions of bisleptons with Higgsino triplet are given
in terms of the mass eigenstates in Eqs. (\ref{la}) and (\ref{nub}) by
\begin{eqnarray}
{\cal L}_{\tilde \Delta}
&=& - {\sqrt 2} {\cal F}^{\tilde l}_{ia}
{\bar l}_i^c P_L {\tilde \Delta}^{++} {\tilde l}_a
- {\cal F}^{\tilde \nu}_{ib} {\bar l}_i^c
P_L {\tilde \Delta}^+ {\tilde \nu}_b
\nonumber \\
&& - {\cal F}^{\tilde l}_{ia} {\bar \nu}_i^c
P_L {\tilde \Delta}^+ {\tilde l}_a
+ {\sqrt 2} {\cal F}^{\tilde \nu}_{ib}
{\bar \nu}_i^c P_L {\tilde \Delta}^0 {\tilde \nu}_b
\nonumber \\
&& + {\rm H.c.} ,
\label{susy-f}
\end{eqnarray}
where
\begin{eqnarray}
{\cal F}^{\tilde l}_{ia} = ( f U^{\tilde l \dagger} )_{ia} , \
{\cal F}^{\tilde \nu}_{ib} = ( f U^{\tilde \nu \dagger} )_{ib} .
\end{eqnarray}

The contributions of $ L $-$ \Delta $ loops are calculated
by using the interactions in Eq. (\ref{nonsusy-f}) as
\begin{eqnarray}
A_R^{\Delta} & = & \frac{1}{32 \pi^2} \frac{ m_{l_j} }{m_\Delta^2}
\sum_k f_{ik}^* f_{jk} \left[ F_1 (0) + 4 F_1 (x_k) - 2 F_2 (x_k) \right] ,
\nonumber \\ \\
A_L^{\Delta} & = & ( m_{l_i} / m_{l_j} ) A_R^{\Delta} ,
\end{eqnarray}
where $ x_k \equiv ( m_{l_k} / m_\Delta )^2 $
with the scalar Higgs triplet mass $ m_\Delta $ in Eq. (\ref{mDelta}).
The functions $ F_1 $ and $ F_2 $ are given by
\begin{eqnarray}
F_1 (x) &=& \frac{1 - 6 x + 3 x^2 + 2 x^3 - 6 x^2 \ln x}{ 6 (1-x)^4 },
\\
F_2 (x) &=& \frac{ 2 + 3 x - 6 x^2 + x^3 + 6 x \ln x }{ 6 (1-x)^4 } .
\end{eqnarray}
The contributions of $ {\tilde L} $-$ {\tilde \Delta} $ loops
are also calculated by using the interactions in Eq. (\ref{susy-f}) as
\begin{eqnarray}
A_R^{\tilde \Delta}
& = & \frac{1}{32 \pi^2} \frac{ m_{l_j} }{M_\Delta^2}
\left\{ \sum_a {\cal F}^{\tilde l *}_{ia} {\cal F}^{\tilde l}_{ja}
\left[ - 2 F_1 (x_a) + 4 F_2 (x_a) \right] \right.
\nonumber \\
&& \left. +
\sum_b {\cal F}^{\tilde \nu *}_{ib} {\cal F}^{\tilde \nu}_{jb} F_2 (x_b)
\right\} ,
\\
A_L^{\tilde \Delta}
& = & ( m_{l_i} / m_{ l_j } ) A_R^{\tilde \Delta} ,
\end{eqnarray}
where $ x_a \equiv ( M_{{\tilde l}_a} / M_\Delta )^2 $ and
$ x_b \equiv ( M_{{\tilde \nu}_b} / M_\Delta )^2 $,
and the Higgsino triplet mass is given by
$ M_{\tilde \Delta} = M_\Delta $.

Here, two remarks should be made.
(i) The suppression factor $ m_{l_i} / m_{l_j} \ll 1 $ appears
in the left-handed contributions
where the chirality is flipped in the final state $ l_i $.
This is due to the fact that only the left-handed lepton doublets
participate in the $ f $ coupling of bileptons and Higgs triplet.
(ii) These amplitudes are approximately proportional to
$ ( f^\dagger f )_{ij} $.
In the amplitudes $ A^{\Delta}_{L,R} $
we have $ F_{1,2} ( x_k ) \simeq F_{1,2} (0) $ for $ x_k \ll 1 $,
so that the factor $ ( f^\dagger f )_{ij} = \sum_k f_{ik}^* f_{jk} $
($ f = f^{\rm T} $) is extracted.
Similarly, in the amplitudes $ A^{\tilde{\Delta}}_{L,R} $
we may neglect the mass differences among the sleptons
for small enough $ | f | \lesssim 0.1 $,
so that the factor $ ( f^\dagger f )_{ij}
= \sum_a {\cal F}^{\tilde l *}_{ia} {\cal F}^{\tilde l}_{ja}
= \sum_b {\cal F}^{\tilde \nu *}_{ib} {\cal F}^{\tilde \nu}_{jb} $
is extracted again
with unitarity of $ U^{\tilde l} $ and $ U^{\tilde \nu} $.
In other words, the flavor mixing of the intermediate sleptons
is actually ineffective for $ A^{\tilde{\Delta}}_{L,R} $
in the leading order of $ | f |^2 $.

\end{document}